\documentclass[a4paper,fleqn,usenatbib]{mnras}

\usepackage{newtxtext,newtxmath}

\usepackage[T1]{fontenc}
\usepackage{ae,aecompl}

\usepackage{multirow}
\usepackage{multicol}  
\usepackage{threeparttable}
\usepackage{subfigure}

\usepackage{graphicx}	
\usepackage{amsmath}	
\usepackage{amssymb}	

\newcommand{\newton}{{\it XMM-Newton}}

\newcommand{\msun}{\ensuremath{M_{\sun}}}

\newcommand{\ha}{H\ensuremath{\alpha}}
\newcommand{\hb}{H\ensuremath{\beta}}
\newcommand{\hc}{H\ensuremath{\gamma}}

\newcommand{\heii}{He\,{\footnotesize II}}
\newcommand{\hei}{He\,{\footnotesize I}}

\newcommand{\oiii}{[O\,{\footnotesize III}]}
\newcommand{\feii}{{\rm Fe\,{\footnotesize II}}}
\newcommand{\sii}{[S\,{\footnotesize II}]}
\newcommand{\nii}{[N\,{\footnotesize II}]}

\title[SDSS J2118$-$0732: a new $\gamma$-ray flaring NLS1]{SDSS J211852.96$-$073227.5: a new $\gamma$-ray flaring narrow-line Seyfert 1 galaxy}

\author[Hui Yang et al.]{Hui Yang,$^{1,2}$\thanks{E-mail: huiyang@nao.cas.cn (HY)}
Weimin Yuan,$^{1,2}$\thanks{E-mail: wmy@nao.cas.cn (WY)}
Su Yao,$^{3,4}$ Ye Li,$^{3,4}$ Jin Zhang,$^1$ Hongyan Zhou,$^{5,6}$
\newauthor S. Komossa,$^7$
He-Yang Liu$^{1,2}$ and Chichuan Jin$^{1}$
\\
$^{1}$Key Laboratory of Space Astronomy and Technology, National Astronomical
Observatories, Chinese Academy of Sciences, \\Beijing 100012, China\\
$^{2}$School of Astronomy and Space Science, University of Chinese Academy of
Sciences, 19A Yuquan Road, Beijing 100049, China\\
$^{3}$Kavli Institute for Astronomy and Astrophysics, Peking University,
Beijing 100871, China\\
$^{4}$National Astronomical Observatories, Chinese Academy of Sciences, Beijing 100012, China\\
$^{5}$Polar Research Institute of China, 451 Jinqiao Road, Shanghai 200136, China\\
$^{6}$University of Science and Technology of China, Chinese Academy of Sciences, Hefei, Anhui 230026, China\\
$^{7}$Max-Planck-Institut f\"ur Radioastronomie, Auf dem H\"ugel 69, D-53121 Bonn, Germany
}

\date{Accepted 2018 April 10. Received 2018 April 2; in original form 2018 January 11}

\pubyear{2018}

\begin{document}
\label{firstpage}
\pagerange{\pageref{firstpage}--\pageref{lastpage}}
\maketitle

\begin{abstract}
We report on the identification of a new $\gamma$-ray-emitting narrow-line Seyfert 1 (NLS1)
galaxy, SDSS J211852.96$-$073227.5 (hereinafter J2118$-$0732).
The galaxy, at a redshift of 0.26, is associated with a radio source of flat/inverted spectrum at high radio frequencies.
The analysis of its optical spectrum obtained in the Sloan Digital Sky Survey (SDSS) revealed a small linewidth of the broad component of the \hb\ line (full width at half-maximum = 1585 km\,s$^{-1}$), making it a radio-loud NLS1 galaxy -- an intriguing class of active galactic nuclei with exceptional multiwavelength properties.
A new $\gamma$-ray source centred at J2118$-$0732 was sporadically detected during 2009--2013 in form of flares by the {\it Fermi}-LAT.
Our \newton\ observations revealed a flat X-ray spectrum described by a simple power law,
and a flux variation by a factor of $\sim$2.5 in five months. 
The source also shows intraday variability in the infrared band.
Its broad-band spectral energy distribution can be modelled by emission from a simple one-zone leptonic jet model, and the flux drop from infrared to X-rays in five months can be explained by changes of the jet parameters, though the exact values may be subject to relatively large uncertainties.
With the NLS1-blazar composite nucleus, the clear detection of the host galaxy, and
the synchronous variations in the multiwavelength fluxes,
J2118$-$0732 provides a new perspective on the formation and evolution of relativistic jets
under the regime of relatively small black hole masses and high accretion rates.
\end{abstract}




\section{Introduction}

A new class of $\gamma$-ray loud active galactic nuclei (AGNs) has been emerging
since the first detection of $\gamma$-ray emission from the radio-loud (RL)
narrow-line Seyfert 1 galaxies (NLS1s) by the Large Area Telescope (LAT) onboard the {\it Fermi} $\gamma$-ray Space Telescope
(hereinafter, {\it Fermi}) satellite \citep{2009ApJ...699..976A}.
Distinctly different from the well-known paradigm that powerful relativistic jets are generally
associated with elliptical galaxies in typical RL AGNs like blazars and radio galaxies,
$\gamma$-ray emitting NLS1s have drawn great attention from the AGN community
in the past decade.

NLS1s are conventionally defined as type 1 AGNs with 
the full width at half-maximum (FWHM) of the broad \hb\ line less than 2000\,km\,s$^{-1}$,
weak \oiii\ emission (\oiii/\hb\,< 3), and usually strong \feii\ emission lines
\citep[][]{1985ApJ...297..166O}.
Several correlations have been found among optical emission lines and X-ray properties
referred to as the eigenvector 1 (EV1) correlations
\citep{1992ApJS...80..109B,2002ApJ...565...78B}. 
Given their relatively small widths of the broad lines, NLS1s tend to have lower
black hole (BH) masses and higher Eddington ratios near or above their Eddington limits
\citep{2004AJ....127.3168B,2004ApJ...606L..41G,2012AJ....143...83X,2015AJ....150...23Y,2017MNRAS.468.3663J,2017MNRAS.471..706J}.
However, we note in passing that it has been occasionally suggested that
BH masses in NLS1s could be higher \citep[e.g.][]{2013MNRAS.431..210C,2016MNRAS.458L..69B,2017MNRAS.469L..11D}
while other studies continue to favour low BH masses \citep[e.g.][]{2009ApJ...707L.142A,2015AJ....150...23Y,2016MNRAS.463.4469D,2017MNRAS.464.2565L}.
Furthermore, NLS1s are found to preferably be hosted in disc-like galaxies with pseudo-bulges
\citep{2006AJ....132..321D,2012ApJ...754..146M}.

Evidence that indicates the presence of relativistic jets in some
RL NLS1s has been accumulated in recent years \citep{2006AJ....132..531K}, especially at the highest radio loudness regime,
due to their blazar-like characteristics \citep{2007ApJ...658L..13Z,2008ApJ...685..801Y}.
RL NLS1 usually shows a compact radio morphology with a one-sided core-jet structure
detected at parsec scale \citep[][]{2015ApJS..221....3G}.
In some objects, the radio emission extends up to kpc scales \citep[e.g.][]{2008A&A...490..583A,2012ApJ...760...41D,2015ApJ...800L...8R}.
A significant fraction of RL NLS1s presents a flat/inverted radio spectrum,
usually with a very high brightness temperature \citep{2008ApJ...685..801Y,2015ApJS..221....3G}.
Some RL NLS1s are found to have intraday infrared variability
\citep[e.g.][]{2012ApJ...759L..31J,2015MNRAS.454L..16Y}.
Besides, superluminal motions have also been found in some RL NLS1s
(e.g.  SBS~0846$+$513: \citealt{2013MNRAS.436..191D};
1H~0323$+$342: \citealt{2016RAA....16..176F}).
Since the launch of the {\it Fermi} satellite in 2008, $\gamma$-ray detections from some
genuine RL NLS1s have confirmed the existence of relativistic jets in this new class of
$\gamma$-ray loud AGNs \citep{2009ApJ...699..976A,2009ApJ...707L.142A,2011nlsg.confE....F,2012MNRAS.426..317D,2015MNRAS.452..520D,2015MNRAS.454L..16Y,2018ApJ...853L...2P}.
With their extreme distributions in AGN parameter space and different host environments,
RL NLS1s allow us to readdress some of the key questions regarding the formation
and evolution of relativistic jets under extreme physical conditions
as well as the coupling of jets and accretion flows.
However, further investigation is hindered by the scarcity of this particular class of objects.

In this paper, we report on the discovery of a new $\gamma$-ray source detected by
the {\it Fermi}-LAT associated with a RL NLS1, SDSS J211852.96$-$073227.5 (hereinafter J2118$-$0732).
This $\gamma$-ray-emitting NLS1 was found in our ongoing study on a sample of RL NLS1s
\citep{2007ApJ...658L..13Z,2008ApJ...685..801Y,2015A&A...574A.121K,2015MNRAS.454L..16Y}.
While we were analysing this NLS1, it was independently reported as
a new $\gamma$-ray source by \citet{2018ApJ...853L...2P} and
in The Preliminary {\it Fermi}-LAT 8-yr Point Source List (FL8Y)\footnote{\url{https://fermi.gsfc.nasa.gov/ssc/data/access/lat/fl8y/}},
and classified as an NLS1 by \citet{2017ApJS..229...39R}.
It was also included in the Combined Radio All-Sky Targeted Eight GHz Survey
\citep[CRATES;][]{2007ApJS..171...61H} catalogue
as a flat-spectrum radio source due to its flat spectrum below 4.8 GHz.
The object was detected in the ROSAT All Sky Survey (RASS), with the total counts
$<$ 20 \citep{2007AJ....133..313A}.
The X-ray spectral and timing properties are essential for linking the postulated high energy
$\gamma$-ray emission with other wavelength data of this AGN,
and for understanding its radiation mechanism.
To this end, we proposed to observe J2118$-$0732 with the \newton\ at two epochs
separated by about five months to study its X-ray spectral and timing properties.
We report the results of these observations in this paper.

The paper is organized as follows.
In Section~\ref{sec:spectroscopy}, we report the optical spectroscopic analysis and results.
The {\it Fermi}-LAT data selection and analysis are included in Section~\ref{sec:fermi}.
We report the \newton\ observations and results in Section~\ref{sec:xmm}.
The radio properties, infrared variability, and optical/ultraviolet (UV) characteristics are discussed
in Section~\ref{sec:radio}, Section~\ref{sec:wise}, and Section~\ref{sec:opt/uv}, respectively.
The broad-band spectral energy distribution (SED) modelling is given in Section~\ref{sec:sed},
followed by discussions in Section~\ref{sec:discussion} and a summary in Section~\ref{sec:summary}. 
Throughout this paper, we assume a cosmology with
$H_0$ = 67.8 km\,s$^{-1}$\,Mpc$^{-1}$, $\Omega_\Lambda$ = 0.692,
and $\Omega_m$ = 0.308 \citep[][]{2016A&A...594A..13P}.

\section{Optical Spectroscopy and the NLS1 Classification}

\label{sec:spectroscopy} 

\begin{figure}
	\includegraphics[width=\columnwidth]{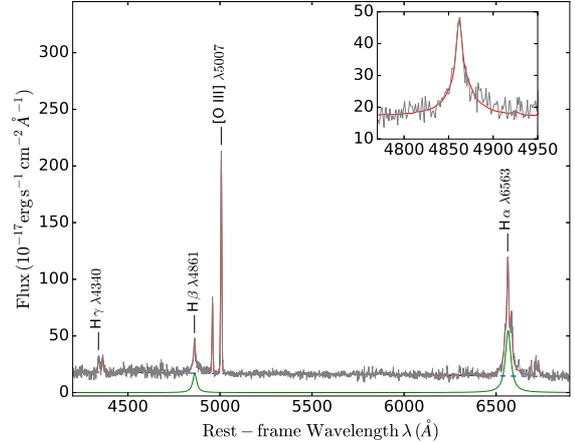}
    \caption{The rest-frame SDSS spectrum of J2118$-$0732 after correction
    for Galactic extinction. The solid red line shows the best-fitting model of total fluxes.
    The dashed blue line represents a broken power-law continuum and the solid green line below the continuum accounts for the broad components of the \ha\ and \hb\ lines.
    The zoom-in \hb\ line is displayed in the inset panel.}
    \label{fig:sdss_spectrum}
\end{figure}

The optical spectrum was obtained by the Sloan Digital Sky Survey (SDSS) on 2001 August 25, with an exposure time of 911\,s. 
After being corrected for Galactic extinction with $E(B-V)$=0.18\,mag
\citep[][]{2011ApJ...737..103S} and an $R_V$= 3.1 extinction law,
the spectrum is transformed into the rest frame with a redshift of $z$=0.26
(see Fig.~\ref{fig:sdss_spectrum}).

In the spectrum, the host galaxy contribution is negligible.
Thus, a similar strategy as in \citet{2008MNRAS.383..581D} and \citet{2015MNRAS.454L..16Y}
was adopted to fit the spectrum. 
We fitted simultaneously the continuum, the \feii\ multiplets, and other emission lines
in the range of 4200--6900\,\AA.
A broken power law (PL) with a break wavelength of 5600\,\AA\ and
the optical \feii\ emission multiplets modelled with the templates from
\citet{2008MNRAS.383..581D} were used to fit the so-called pseudo-continuum.
The emission lines identified from the composite SDSS quasar spectrum
\citep[see table 2 in][]{2001AJ....122..549V} from \hc\ to 
\sii\,$\lambda\lambda6716,6731$\footnote{\label{foot:lines}
Several emission lines were masked out, for either they were too weak to constrain in the fit
or had little effect on the results. \hc\,$\lambda$4340, \oiii\,$\lambda$4363,
\heii\,$\lambda4686$, \hb\,$\lambda4861$, \oiii\,$\lambda\lambda4959,5007$,
\hei\,$\lambda4861$, \nii\,$\lambda \lambda6548,6583$, \ha\,$ \lambda6563$,
and \sii\,$\lambda\lambda6716,6731$ were included in our spectroscopic analysis.}
were modelled as follows.
The broad \ha\ and \hb\ were assumed to have the same redshift and profile fitted by
either a single Lorentzian or concentric double Gaussians.
Other broad emission lines were modelled with one Gaussian.
One single Gaussian with a constraint of FWHM < 1000 km\,s$^{-1}$ was used
for modelling each narrow emission line. 
Both flux ratios of the \nii\ doublets and the \oiii\ doublets
were fixed to their theoretical values of 3.

\begin{table}
	\caption{The spectroscopic fitting results and basic parameters of J2118$-$0732.}
	\label{tab:spectroscopic}
   \begin{threeparttable}
	\begin{tabular}{lrr} 
	   \hline
		Model$^a$  & L &  2G       \\
		\hline
		FWHM(\hb)/km\,s$^{-1}$     &  $1585 \pm 182$  & $1882 \pm 375$    \\
		FWHM(\oiii)/km\,s$^{-1}$     &  $364 \pm 15$    & $327 \pm 14$    \\
		\oiii/\hb\  &		 1.7    & 2.0  \\
       $R_{4570}$ & 0.06 & 0.06 \\
       $M_{\rm BH}$/10$^7$\,\msun\ & 3.4 & 3.7 \\
      Eddington ratio & 0.15 & 0.11\\
     \hline
	\end{tabular}
	\begin{tablenotes}
	\item[]
     $^a$ Models used for fitting the broad \hb\ emission lines are: 
     one Lorentzian profile (L) or a double-Gaussian profile (2G).\\
     \end{tablenotes}
	\end{threeparttable}
\end{table}

The broad \hb\ line was equally well fitted with either a Lorentzian profile or
a concentric double-Gaussian profile. 
In the former case, the width of the broad component is
FWHM(H$\beta_{\rm broad}$) = 1585 $\pm$ 182\,km\,s$^{-1}$, while in the latter case it is
FWHM(H$\beta_{\rm broad}$) = 1882 $\pm$ 375\,km\,s$^{-1}$
after subtracting the effect of instrumental broadening.
\citet{2011ApJS..194...45S} also obtained a similar
FWHM(H$\beta_{\rm broad}$) = 1597 $\pm$ 441\,km\,s$^{-1}$
using a single Gaussian or multiple Gaussians model.
In the following, we adopted the fitting results using the Lorentzian profile
since the Lorentzian model is better suited for the broad-line profiles of NLS1s
\citep[e.g.][]{2001A&A...372..730V,2006ApJS..166..128Z}.
The line width of \oiii\,$\lambda5007$ is 364 $\pm$ 15\,km\,s$^{-1}$
and there is no obvious evidence of blue wings in \oiii\ lines.
The flux ratio of \oiii\,$\lambda5007$ to H$\beta_{\rm total}$ is $\approx1.7$
and $R_{4570} \equiv$ \feii\,$\lambda4570$/H$\beta_{\rm total}\approx0.06$
(\feii\ is calculated by integrating from 4434 to 4684\,\AA).
The small broad-line width and the relatively weak \oiii\ of J2118$-$0732
fulfil the conventional definition of an NLS1.
However, the \feii\ emission is rather weak compared to most NLS1s
and the optical continuum is red with
$\alpha_{\rm opt}$ = -1.5 ($S_{\nu}\propto \nu^{\alpha}$) for J2118$-$0732.
Nevertheless, there is no quantitative definition for the \feii\ strength,
and some NLS1s do emit weak \feii\ emission
\citep{2006ApJS..166..128Z,2012AJ....143...83X,2016MNRAS.462.1256C,2017ApJS..229...39R}.

Under the assumption of a virialized broad-line region (BLR), we can estimate
the BH mass of J2118$-$0732 using FWHM(H$\beta_{\rm broad}$) = 1585\,km\,s$^{-1}$
and the \hb-based estimator from \citet{2009ApJ...707.1334W},
with a 1$\sigma$ uncertainty of 0.5 dex.
Considering the contamination from the jet contribution to the continuum, a surrogate for the
5100\,\AA\ luminosity ($\lambda  L_{5100}$) estimated from the broad \hb\ line luminosity
using equation (5) from \citet{2006ApJS..166..128Z} was applied.
We found $M_{\rm BH} \approx 3.4 \times 10^7\,M_{\sun}$
($M_{\rm BH} \approx 3.7 \times 10^7\,M_{\sun}$ if the line parameters fitted from
the double-Gaussian profile were used), which is consistent with the estimations from
\citet{2007ApJ...667..131G} and \citet{2008ApJ...688..826L} based on
other commonly used single-epoch formalisms.

The bolometric luminosity can be calculated assuming $L_{\rm bol} = 9\lambda L_{5100}$
(estimated from \hb), as suggested by \citet{2000ApJ...533..631K},
which results in $L_{\rm bol} \approx 6.2 \times 10^{44}$\,erg\,s$^{-1}$.
The Eddington ratio of this object is $\lambda \approx 0.15$,
relatively low compared to other NLS1s but higher than typical broad-line Seyfert 1 galaxies
\citep{2012AJ....143...83X,2016MNRAS.462.1256C,2017ApJS..229...39R}.
The bolometric luminosity and the Eddington ratio of J2118$-$0732 estimated by \citet[][]{2011ApJS..194...45S} are
$L_{\rm bol} \approx 1.1 \times 10^{45}$\,erg\,s$^{-1}$ and $\lambda \approx 0.39$,
slightly larger than our estimations. 
This discrepancy has mainly resulted from the overestimation of the bolometric luminosity
using the directly measured luminosity at 5100~\AA,
where the jet contribution cannot be ignored, especially for those RL AGNs.
The fitting results and the basic parameters estimated for J2118$-$0732 are summarized
in Table~\ref{tab:spectroscopic}.

\section{$\gamma$-ray data analysis}
\label{sec:fermi} 

\subsection{Observations and data reduction}

The {\it Fermi}-LAT is a pair-conversion $\gamma$-ray telescope operating
from 20\,MeV to > 300\,GeV.
It has a large peak effective area ($\sim$8000\,cm$^2$ for 1\,GeV photons),
allowing a large field of view ($\sim$2.4\,sr) with an angular resolution
(68 per cent containment radius) better than 1\degr\ for energies above 1\,GeV
and an energy resolution of typically $\sim$10 per cent. 
Further details about the {\it Fermi}-LAT are given in \citet{2009ApJ...697.1071A}.

We analysed the Pass 8 data collected by the LAT from 2008 August 4th (Modified Julian Day, MJD 54 682) to 2017 July 14 (MJD 57 948). 
During this period, the LAT instrument operated in survey mode,
scanning the entire sky every 3\,h. 
The data analysis was performed with the standard \texttt{ScienceTools}
version v10r0p5 software package.
We used the standard binned maximum likelihood method implemented in
the Science tool \texttt{gtlike} to analyse
each time bin longer than or equal to 1\,yr. 
For smaller time bins, we used the unbinned method where the number of events
was expected to be small. 
The LAT data were extracted from a region of interest (ROI) defined as a circle
of 30\degr\ radius centred at the location of J2118$-$0732 for the binned method
and a circle of 20\degr\ radius for the unbinned method.
Only events belonging to the 'Source' class from 100\,MeV to 300\,GeV
were used with \texttt{P8R2\_SOURCE\_V6} Instrument Response Functions.
In addition, a cut on the zenith angle (< 90\degr) was applied to
minimize the contamination from Earth limb $\gamma$-rays.
Isotropic (iso\_P8R2\_SOURCE\_V6\_v06.txt) and Galactic diffuse emission (gll\_iem\_v06.fits) components were used to model the background. 
The normalization of each component was set to be free during the spectral fitting.

The significance of the $\gamma$-ray signal from the source was evaluated by means of a
maximum-likelihood test statistic (TS) that results in TS = 2 (log$L_1$ - log$L_0$),
where $L$ is the likelihood of the data given the model with ($L_1$) or without ($L_0$)
a point source at the position of J2118$-$0732 \citep[e.g.][]{1996ApJ...461..396M}.
Because our data cover a 9-yr period, the source model included
all the point sources from The Third {\it Fermi}-LAT source catalogue \citep[3FGL;][]{2015ApJS..218...23A}
that fall within the ROI and an addition of 10\degr\ annular radius around it,
and new sources detected in FL8Y within 10\degr\ of J2118$-$0732.\footnote{Before
the recent publication of FL8Y, we have discovered the $\gamma$-ray detection from the optical
position of J2118$-$0732, just using 3FGL sources for source model in the $\gamma$-ray data analysis.}
The spectra of these sources were parameterized by a PL, a log-parabola,
or a super exponential cut-off, as in the 3FGL catalogue,
and by a PL for 18 new sources in FL8Y.
Sources within 10\degr\ of J2118$-$0732 and with their significances greater than 5
calculated from 3FGL and FL8Y were included with the normalization factors
and the photon indices left as free parameters.
For other sources in the source model, we fixed the normalizations and the photon indices
to the values in the catalogues.
The uncertainties correspond to 68 per cent confidence for the {\it Fermi}-LAT results.

\subsection{Results}
\label{sec:fermi-results}

\begin{figure}
	\includegraphics[width=\columnwidth]{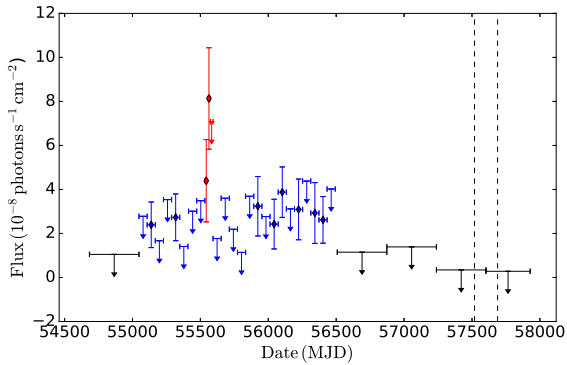}
	\caption{Integrated flux light curve of J2118$-$0732 obtained by the {\it Fermi}-LAT 
	in the 0.1--300\,GeV energy range during 2008 August 4--2017 July 14
	(MJD 54 682--57 948) with 1-yr bins in black, 60-d bins in blue and 20-d bins in red. 
	Diamonds refer to the detections with higher TS values. 
	Arrows refer to 95 per cent upper limits on the source flux. Upper limits are computed when
	TS < 9 for 1-yr bins and TS < 5 for 60-d bins and 20-d bins. The two dashed
	vertical lines correspond to the epochs of our two \newton\ Observations,
	X-2017-05-10 and X-2017-10-27.}
	\label{fig:gamma_lc}
\end{figure}

\begin{table*}
   
	\centering
	\caption{Results of the {\it Fermi}-LAT data analysis.}
	\label{tab:lat}
	\begin{threeparttable}
	\begin{tabular}{lcccccccc} 
	\hline
	           & \multicolumn{2}{c}{Optimized position (J2000)} & Separation from& $R_{95 \%}$ &
	           $F_{0.1-300 {\rm GeV}}$    & $\Gamma_\gamma$ & $L_\gamma$ & TS  \\
	           &   hh mm ss.s      & dd mm ss.s         & optical position (\degr)  &  (\degr)&
	           (10$^{-9}$\,ph\,cm$^{-2}$\,s$^{-1}$) &   & (10$^{45}$\,erg\,s$^{-1}$) & \\
   \hline
        (1)  &                       &					            &           &                 & 
        		6.6 $\pm$ 1.8	   & 2.65 $\pm$ 0.16  & 0.7 $\pm$ 0.3     & 27  \\    
        (2)  &  21 18 45.4    &  -07 25 02.1         &     0.128   &  0.161            &   
             15.4 $\pm$ 3.1   &  2.79 $\pm$ 0.16 & 1.5 $\pm$ 0.4     &  50   \\
        (3)  &  21 18 37.1    & -07 28 42.7          &  0.091 & 0.264                  &  
              43.7 $\pm$ 1.6  &  2.62 $\pm$ 0.01 &  4.6 $\pm$ 0.2    & 24  \\
        (4)  & 21 18 58.9     & -07 30 04.5          &   0.047 &  0.14                    &  
              9.2 $\pm$ 2.0    & 2.80 $\pm$ 0.15 & 0.9 $\pm$ 0.3      & 40  \\
       (5)   & 21 19 14.1     & -07 28 31.1         &  0.109 & 0.20$\times$0.16 &
             13.2 $\pm$ 2.5   & 2.81 $\pm$ 0.12 & 1.3 $\pm$ 0.3       & 47 \\
 \hline
 \end{tabular}
    \begin{tablenotes}
\item[] 
 (1) our results analysed from the whole period during 2008 August 4--2017 July 14. \\
 (2) our results analysed from the active 4 yr during 2009 August 5--2013 July 20. \\
 (3) our results analysed  from the flare during 2010 December 2--2011 February 1. \\
 (4) \citet[][]{2018ApJ...853L...2P} results analysed from the period during 2008 August 5--2017 January 5. \\ 
 (5) FL8Y results analysed from the period during 2008 August 4--2016 August 2. \\
 		\end{tablenotes}
 	\end{threeparttable}
\end{table*}

First, we tested whether a $\gamma$-ray source at the optical position of J2118$-$0732
was detected by the {\it Fermi}-LAT over the whole duration from 2008 August 4 to
2017 July 14 with a binned maximum likelihood analysis using a PL model.
A source was clearly detected with a TS value of 27 ($\sim$5\,$\sigma$) for J2118$-$0732,
an average $\gamma$-ray integrated photon flux from 0.1 to 300\,GeV of
$F_{0.1-300 {\rm GeV}}$ = ($6.6 \pm 1.8$)\,$\times\,10^{-9}$\,ph\,cm$^{-2}$\,s$^{-1}$,
and a photon index $\Gamma_{\gamma}=2.65\pm0.16$.
This flux corresponds to an isotropic $\gamma$-ray luminosity of
$L_\gamma$ = (0.7 $\pm$ 0.3)\,$\times\,10^{45}$\,erg\,s$^{-1}$
in the 0.1--300\,GeV rest-frame energy band.
The isotropic $\gamma$-ray luminosity is given as $L_\gamma$ = $4\pi$\,$d_L^2$\,$S_{\gamma} $,
where $d_L$ is the luminosity distance and $S_{\gamma}$ is the flux density. 

Next, the integrated flux of each year during the whole 9-yr {\it Fermi}-LAT observation
was calculated to investigate the long-term $\gamma$-ray variability of J2118$-$0732.
For each 1-yr bin, the spectral parameters of J2118$-$0732 and all sources within
10\degr\ were frozen to the values obtained from the likelihood analysis over the entire period.
J2118$-$0732 was relatively active during 2009 August 5--2013 July 20 with higher flux
measurements and significant TS values, but remained undetected in other years.
The 95 per cent upper limits were provided if TS < 9 during 2008 August 4--2009 August 5 and 2013 July 20--2017 July 14 with 1-yr bins in the long-term $\gamma$-ray light
curve of J2118$-$0732 in Fig.~\ref{fig:gamma_lc}.
A further likelihood analysis of the data taken during the relatively active 4 yr
resulted in a TS of 50 ($\sim$7\,$\sigma$).
We obtained a photon flux index of $\Gamma_{\gamma}=2.79\pm 0.16$ and a flux of
$F_{0.1-300 {\rm GeV}}$ = (15.4 $\pm$ 3.1)\,$\times\,10^{-9}$\,ph\,cm$^{-2}$\,s$^{-1}$,
corresponding to an isotropic $\gamma$-ray luminosity of
$L_\gamma$ = (1.5 $\pm$ 0.4)\,$\times\,10^{45}$\,erg\,s$^{-1}$ during the 4-yr period.
We performed a $\gamma$-ray point source localization using the \texttt{gtfindsrc} tool
over the photons in the energy band from 1 to 300\,GeV extracted during this period.
The fit resulted in R.A. = 21$^{\rm h}$18$^{\rm m}$45.4$^\mathrm{s}$,
Dec. = $-$7$^{\circ}$ 25$\arcmin$02.1$\arcsec$ with a 95 per cent error circle radius of
$R_{95 \%}$ = 0.161\degr.
Here, we excluded the low-energy photons because of their poor angular resolutions.
The $\gamma$-ray source is located 0.128\degr\ away from the optical position of J2118$-$0732,
which is within the 95 per cent error circle of the $\gamma$-ray position.
 
We also calculated the $\gamma$-ray light curve of J2118$-$0732 during 2009--2013 using
60-d time bins and 20-d time bins and added it into Fig.~\ref{fig:gamma_lc}.
For each time bin, the spectral parameters of J2118$-$0732 and all sources within 10\degr\
were frozen to the values obtained from the 4-yr likelihood analysis.
The 95 per cent upper limit was also calculated if TS $<$ 5.
The systematic uncertainty in the flux is dominated by the systematic uncertainty in the
effective area \citep[][]{2012ApJS..203....4A}, which amounts to 10 per cent at 100\,MeV,
decreasing to 5 per cent at 560\,MeV and increasing to 10 per cent above 10\,GeV.

J2118$-$0732 was detected sporadically by the {\it Fermi}-LAT, with an increase in flux and a high
TS value of 24 ($\sim$5\,$\sigma$) during 2010 December 2--2011 February 1.
The flux was doubled within 20\,d and quickly dropped to a lower state below the
{\it Fermi}-LAT detection limit.
During this flaring period, we obtained a photon index of $\Gamma_{\gamma}$ = 2.62 $\pm$ 0.01 and a flux of
$F_{0.1-300 {\rm GeV}}= (43.7 \pm 1.6)\,\times$\,10$^{-9}$\,ph\,cm$^{-2}$\,s$^{-1}$,
corresponding to an isotropic $\gamma$-ray luminosity of
$L_\gamma$ = (4.6 $\pm$ 0.2)\,$\times$\,10$^{45}$\,erg\,s$^{-1}$,
a factor of $\sim$7 higher than the average value during the whole period.
A point source localization using full energy band photons was also performed during this 60-d high state.
The fit resulted in R.A. = 21$^{\rm h}$18$^{\rm m}$37.1$^{\rm s}$, 
Dec. = $-$7$^{\circ}$28$\arcmin$42.7$\arcsec$ which is 0.091\degr\ away from the
optical position of J2118$-$0732 with a 95 per cent error circle radius of $R_{95 \%}$ = 0.264\degr. 

The $\gamma$-ray detection of this source has also been independently
reported by \citet[][]{2018ApJ...853L...2P} and FL8Y.
A brief comparison between their results and ours regarding the $\gamma$-ray properties
is summarized in Table~\ref{tab:lat}.
The $\gamma$-ray detected source suffers from a large spatial uncertainty given
the limited resolution of the {\it Fermi}-LAT. 
So the association of the $\gamma$-ray source with J2118$-$0732
will be discussed further in Section~\ref{sec:association}.

\section{X-ray observations and data analysis}
\label{sec:xmm}

\subsection{XMM observations and data reduction}

\begin{table}
	\caption{XMM EPIC observations for J2118$-$0732.}
	\label{tab:x_obs}
		\begin{threeparttable}
	\begin{tabular}{lcccc} 
		\hline
		Data Set   &   \multicolumn{2}{c}{X-2016-05-10}      &  \multicolumn{2}{c}{X-2016-10-27} \\
		\hline
	  Observation ID &    \multicolumn{2}{c}{0784090201}   &  \multicolumn{2}{c}{0784090301}  \\
		
	  Start Time    &   \multicolumn{2}{c}{2016 May 10}      & \multicolumn{2}{c}{2016 October 27} \\
		(UTC)          &    \multicolumn{2}{c}{12:36:12}          &      \multicolumn{2}{c}{10:55:03}         \\
		
		Duration (ks) &		 \multicolumn{2}{c}{31.9}  	 & 	  \multicolumn{2}{c}{32.8}      \\
		
		 Detectors          &      PN   & MOS1/2                             &   PN   & MOS1/2  \\
		 				
		Good Exposure  &		18.6 	 & 	29.3  & 16.2 & 29.4       \\
		 (ks)   &   & &  & \\
		  Net Count Rate & 0.193   &  0.117 & 0.069 & 0.043 \\
		(count s$^{-1}$)     &   & &  & \\
		\hline
	\end{tabular}
		\end{threeparttable}
\end{table}

We conducted two observations on J2118$-$0732 with the \newton\ \citep{2001A&A...365L...1J}
in 2016: for 31.9 ks on 2016 May 10 and for 32.8 ks on 2016 October 27
(Observation IDs 0784090201 and 0784090301, PI: Su Yao), hereinafter X-2016-05-10 and X-2016-10-27.
All observations were performed with the European Photon Imaging Camera (EPIC) using
PN \citep[][]{2001A&A...365L..18S} and MOS \citep[][]{2001A&A...365L..27T} CCD arrays
as well as the Optical Monitor \citep[OM;][]{2001A&A...365L..36M}.
OM data analysis will be discussed in Section~\ref{sec:opt/uv}.
All EPIC observations operated in prime large window mode with the thin filter.
The \newton\ EPIC observations are summarized in Table~\ref{tab:x_obs}.
Observation data files were processed to create calibrated event lists and full frame images
following standard procedures with the \newton\ Science Analysis System (\textsc{sas} Version 16).
Source regions were extracted from a circular aperture around J2118$-$0732
with an optimized radius of 35$\arcsec$.
The background regions were chosen from either a source-free circle for PN or a source-centre
annulus for MOS with similar readout distances to ensure similar background noise levels.
A weak X-ray source near J2118$-$0732 positioned at
R.A. = 21$^{\rm h}$18$^{\rm m}$53.76$^{\rm s}$, Dec. = $-$7$^{\circ}$32$\arcmin$2.4$\arcsec$
was detected by the \textsc{sas} algorithm \texttt{edetect\_chain} and was excluded during the region selections.
Pile-up was negligible in all observations and flaring particle background was filtered.
The resulting net (source minus background) PN count rates and good exposures are
0.193\,cts\,s$^{-1}$ and 18.6\,ks (0.117\,cts\,s$^{-1}$ and 29.3\,ks for the combined MOS spectrum)
and 0.069\,cts\,s$^{-1}$ and 16.2\,ks (0.043\,cts\,s$^{-1}$ and 29.4\,ks
for the combined MOS spectrum) for the first and the second observations, respectively. 
The uncertainties correspond to 90 per cent confidence for EPIC results.

\subsection{X-ray spectra and variability}

\begin{figure}
\centering
\includegraphics[width=\columnwidth]{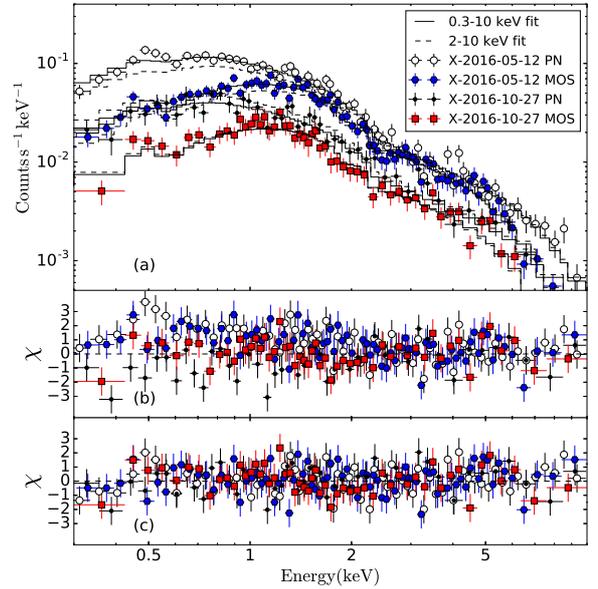}
\caption{The \newton\ EPIC spectra and best-fitting results of a single power-law model modified by 
			  the total Galactic absorption for X-2016-05-10 and X-2016-10-27. 
			  Upper panel (a): X-2016-05-12 PN (white circles), X-2016-05-12 MOS (blue hexagons), 
			  X-2016-10-27 PN (black dots), and X-2016-10-27 MOS data (red squares), 
			  as well as the corresponding best fits for an absorbed PL evaluated over 0.3--10\,keV
			  (black solid line) and 2--10\,keV (black dashed line). Mid panel (b): residuals for fitting the 2--10\,keV
			  energy alone, extrapolated to lower energies. Lower panel (c): residuals for the best fits over 
			  the whole 0.3--10\,keV energy range.}
\label{fig:x_spectra}
\end{figure}

\begin{table}
	\caption{Spectral fits to XMM EPIC observations for J2118$-$0732}
	\label{tab:x_spectra_para}
		\begin{threeparttable}
	\begin{tabular}{lcc} 
		\hline
		Data Set   &  X-2016-05-10 & X-2016-10-27 \\
		Detectors &  PN$+$MOS  & PN$+$MOS \\  
		
		  \hline
		 \multicolumn{3}{l}{\textsc{tbabs}$\times$\textsc{ztbabs}$\times$\textsc{zpow},
		  fixed $N_{\rm H}$ = $N_{\rm H}^{\rm Gal}= 1.15 \times\,10^{21}$\,cm$^{-2}$} \\
		  \multicolumn{3}{l}{ at $z$ = 0, free $N_{\rm H}$ at $z$ = 0.26,
		  fitted over 0.3--10\,keV}  \\
		 $\Gamma_{\rm X}^{a}$ & $1.78^{+0.05}_{-0.03}$   & $1.69^{+0.10}_{-0.09}$ 	\\
		 $N_{\rm H}$/10$^{21}$  & $0^{+0.2}$  & $0.3^{+0.5}_{-0.3}$  \\
		 $\chi^{2}/$ d.o.f.  &  $133.0/150$ & $87.1/97$ \\
		 \hline
		 \multicolumn{3}{l}{\textsc{tbabs}$\times$\textsc{zpow},
		 fixed $N_{\rm H}$ = $N_{\rm H}^{\rm Gal}= 1.15 \times\,10^{21}$\,cm$^{-2}$ at $z$ = 0,}  \\
		 \multicolumn{3}{l}{fitted over 2--10\,keV and extrapolated to 0.3--10\,keV} \\
		 $\Gamma_{\rm X}^{a}$ & $1.63^{+0.12}_{-0.11}$   & $1.70 \pm 0.19$ 	\\
		 $\chi^{2}/$ d.o.f.  &  $62.2/73$ & $31.5/34$ \\
		 \hline	 
		  \multicolumn{3}{l}{\textsc{tbabs}$\times$\textsc{zpow}, 
		  fixed $N_{\rm H}$= $N_{\rm H}^{\rm Gal}= 1.15 \times\,10^{21}$\,cm$^{-2}$ at $z$ = 0,} \\
		\multicolumn{3}{l}{fitted over 0.3--10\,keV} \\
		 $\Gamma_{\rm X}^{a}$ & $1.78\pm 0.03$   & $1.66 \pm 0.06$ 	\\	
		 Flux$^b$  & 7.55 $\pm$ 0.25 & 2.97 $\pm$ 0.19  \\	
		$L_{\rm X}^c$  & 15.8 $\pm$ 0.6 & 6.1 $\pm$ 0.5  \\	
		 $\chi^{2}/$ d.o.f.  &  $137.5/154$ & $94.4/101$ \\
		 \hline
	\end{tabular}
		\begin{tablenotes}
		\item[]
		$^a$ zpower-law photon index; $^b$ unabsorbed flux integrated from 0.3 to 10\,keV in units of
		10$^{-13}$\,erg\,cm\,$^{-2}$\,s$^{-1}$; $^c$ isotropic X-ray luminosity in the 0.3 -- 10\,keV
		rest-frame energy band in units of 10$^{43}$\,erg\,s$^{-1}$. 
		\end{tablenotes}
		\end{threeparttable}
\end{table}

The MOS1 and MOS2 spectra were co-added to produce a single combined spectrum.
And all the spectra were binned to contain at least 25 counts per bin needed for $\chi^2$ analysis.
We used \textsc{xspec} (v.12.9.1) to fit PN and the combined MOS spectra simultaneously
over the 0.3--10\,keV energy range.

We first adopted a single PL modified by the Galactic hydrogen absorption
plus a free hydrogen column at the redshift of the source using the Tuebingen-Boulder model
(\textsc{tbabs} and \textsc{ztbabs} in \textsc{xspec}).
We also assumed element abundances from \citet[][]{2000ApJ...542..914W} and photoelectric cross-sections from \citet[][]{1996ApJ...465..487V}.
The Galactic hydrogen column was fixed to the total (H I plus H$_2$ ) Galactic column $N_{\rm H}^{\rm Gal}$
of 1.15 $\times \, 10^{21}$\,cm$^{-2}$ from \citet[][]{2013MNRAS.431..394W}.
The intrinsic absorption is negligible as it was fitted close to zero.
So, the absorption was fixed at the Galactic value to reduce the number of free parameters.
We checked for the existence of a soft excess by modelling the spectra only above 2\,keV and
extrapolating the best fit down to 0.3\,keV.
No significant soft X-ray excess emission was observed in both observations [see panel (b) in Fig.~\ref{fig:x_spectra}].
The X-ray spectra were then fitted with a single PL modified by the total Galactic absorption which resulted in good fits.
The best-fitting indices are $\Gamma_{\rm X}$ = $1.78\pm 0.03$ and
$\Gamma_{\rm X}$ = $1.66\pm 0.06$ for X-2016-05-10 and  X-2016-10-27, respectively, which are harder than
normal radio-quiet NLS1s \citep[$\Gamma_{\rm X} \sim$2--4; e.g. ][]{2010ApJS..187...64G,2011ApJ...727...31A},
but similar to some other $\gamma$-ray-emitting NLS1s \citep{2009ApJ...699..976A,2013MNRAS.436..191D}. 
The fact that the X-ray spectrum is well modelled by a hard PL suggests that it is typical
of the X-ray emission from jets in blazar-like AGNs and any contribution from a possible corona may be small.
The spectral fit results reported above are shown in Table~\ref{tab:x_spectra_para} and the best-fitting spectra
of a single power-law model modified by the total Galactic absorption are shown in Fig.~\ref{fig:x_spectra}. 

The X-ray flux in the 0.3--10\,keV band decreased by a factor of  $\sim$2.5 between the two observations. 
However, no significant variability was found within each observation.

\section{Other data}

\subsection{Radio properties}
\label{sec:radio}

\begin{figure}
	\includegraphics[width=\columnwidth]{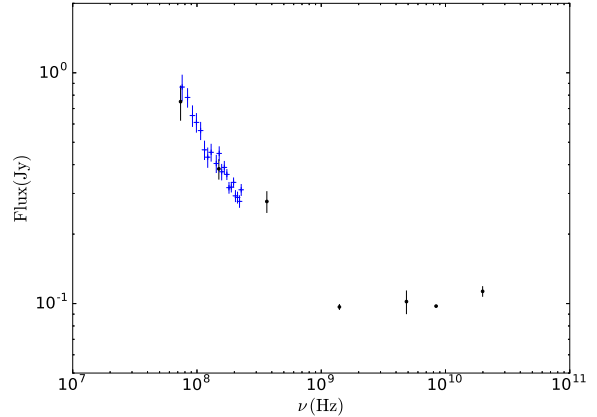}
	\caption{The radio spectrum of J2118$-$0732 collected from NED and VizieR. The simultaneous data across 72$-$231\,MHz indicated with blue cross marks were obtained from the GLEAM survey.}
	\label{fig:radio_spectrum}
\end{figure}

\begin{table*}

	\caption{The GHz radio data of J2118$-$0732 retrieved from NED and VizieR.}
	\label{tab:radio_spec}
	\begin{threeparttable}
	\begin{tabular}{lccccc} 

	\hline
		 Frequency  & Flux   & Polarized & Resolution  &  Observation  & Surveys and     \\
		   (GHz)      &  (mJy)  &   flux (mJy)   &(arc sec) &   date  & references \\
		\hline
		0.074   &  $750 \pm 130 $   &   & 80 & 2003-09-20& (1)      \\
	
       0.15   &  $ 383.5\pm 39$   & & 25 & 2016-03-15 &  (2)      \\
		
	0.17--0.231 & $307.3 \pm 9.1^a$ &  & $\sim100$ & 2013-08-13 & (3) \\
	
		0.365&		 $277 \pm 30$ & &  6  & & (4)   \\
		
		1.4&		 $ 98.7 \pm 20 $ & &  5  & 1997-05-26 &  (5)   \\
		
		1.4&		 $ 96.1 \pm 2.9$ &3.16 &  45  &1993-09-20 &  (6)   \\
		
		4.85 &		 $102 \pm 12 $ &  & 168 &1990-11-06 & (7)     \\
		
		8.4  &		 $97.5 $ &   &  & & (8)    \\
	
	   19.9  &		 $113 \pm 6 $ &  6 & 34  & 2007 Oct 26-30 & (9)    \\
		\hline	
	\end{tabular}
		\begin{tablenotes}
		\item[] Surveys and References: 
		(1) The VLA Low-Frequency Sky Survey \citep[VLSS,][]{2007AJ....134.1245C}, 
		(2) The Giant Metrewave Radio Telescope (GMRT) 150 MHz All-sky Radio Survey 
		\citep[][]{2017A&A...598A..78I}, (3) GaLactic and Extragalactic All-sky Murchison Widefield Array 
		(GLEAM) survey \citep[][]{2017MNRAS.464.1146H},
		(4) The Texas Survey of radio sources at 365MHz \citep[][]{1996AJ....111.1945D}, 
		(5) The Faint Images of the Radio Sky at 20 cm Survey \citep[FIRST,][]{1997ApJ...475..479W}, 
		(6) The NRAO VLA Sky Survey \citep[NVSS,][]{1998AJ....115.1693C}, 
		(7) The Parkes-MIT-NRAO Surveys \citep[PMN surveys,][]{1995ApJS...97..347G}, 
		(8) The Combined Radio All-Sky Targeted Eight GHz Survey \citep[CRATES,][]{2007ApJS..171...61H}, 
		(9) The Australia Telescope 20 GHz Survey \citep[AT20G survey,][]{2010MNRAS.402.2403M} \\
		$^a$ integrated flux between 170 and 231 MHz.
		\end{tablenotes}
		\end{threeparttable}
\end{table*}

J2118$-$0732 was detected as a single unresolved source within 5$\arcsec$ of the SDSS position in several surveys
from the NASA/IPAC Extragalactic Database website (NED)\footnote{\url{http://ned.ipac.caltech.edu}} and
VizieR website\footnote{\url{http://vizier.u-strasbg.fr/index.gml}} listed in Table~\ref{tab:radio_spec}.
The Faint Images of the Radio Sky at 20 cm Survey \citep[FIRST;][]{1997ApJ...475..479W} observation
indicates a compact radio structure at FIRST resolution (5$\arcsec$).
The NRAO VLA Sky Survey \citep[NVSS;][]{1998AJ....115.1693C} 1.4 GHz flux density is 
$96.1 \pm 2.9$\,mJy, corresponding to a radio power of
$P_{\rm 1.4GHz} \sim (1.98 \pm 0.06)\,\times 10^{25}$\,W\,Hz$^{-1}$, and the polarized flux density is
$3.16 \pm 0.49$\,mJy, thus a fractional polarization of $\sim3.3$ per cent. 
The source shows a steep spectrum between 74\,MHz and 1.4\,GHz with
$\alpha_{\rm rad} = -0.7$ \citep[][]{2010A&A...511A..53V}, however,
a flat/inverted spectrum above 1.4\,GHz (see Fig.~\ref{fig:radio_spectrum}).
We refitted the radio spectrum using a broken PL with a break frequency at 1.4\,GHz and obtained
$\alpha_\mathrm{rad} = -0.66$ in the lower frequency band and  $\alpha_{\rm rad} = 0.15$ in the higher
frequency band without considering the variability of the radio flux as the data were not all simultaneous.
However, the synchronous observations across 72--231\,MHz from the GaLactic and Extragalactic All-sky Murchison Widefield Array (GLEAM) survey \citep[][]{2017MNRAS.464.1146H} indicate
the sincerity of a steep radio spectrum at lower frequency.

To estimate the radio loudness parameter defined as
$R_{5 \,{\rm GHz}} \equiv f_{\nu}(5\,{\rm GHz})/f_\nu$(4400\,\AA), 
we used the 4.85\,GHz flux from the Parkes-MIT-NRAO surveys \citep[PMN surveys;][]{1995ApJS...97..347G}
and the 4400 \AA\ flux from the SDSS g band point-spread function (PSF)-model magnitude with Galactic extinction considered.
The $k$-correction with the radio spectral index $\alpha_{\rm rad} = 0.15 $ and
the optical index $\alpha_{\rm opt} = -1.5$ was adopted.
It results in $R_{5\, {\rm GHz}} \sim920$, which is consistent with
$R = f_\nu(5\,{\rm GHz})/f_\nu$(2500\,\AA) $\approx1227$ derived from \citet{2011ApJS..194...45S}. 
However, it cannot be ruled out, though unlikely,
that a strong starburst may contribute to the radio emission.
Such an estimation of radio luminosity from a starburst contribution can be obtained
using the observed far-infrared (FIR)-radio luminosity correlation from \citet[][]{2001ApJ...554..803Y} and
the extrapolated FIR luminosity [using the Wide-field Infrared Survey Explorer \citep[{\it WISE};][]{2010AJ....140.1868W} data and a power-law assumption in the infrared band].
This gives $P_{1.4\,{\rm GHz}} \sim 8.9 \times\,10^{22}$\,W\,Hz$^{-1}$, which is about two orders of magnitude
smaller than the observed radio luminosity, excluding a strong starburst contribution to the radio emission.
Besides, there is also a possibility that intrinsic reddening may affect the optical continuum slope.
Nevertheless, just using the radio power itself, $P_{1.4\, {\rm GHz}} \sim 2\times 10^{25}$\,W\,Hz$^{-1}$, makes J2118$-$0732 formally radio-loud.

\subsection{{\it \textbf{WISE}} infrared data and intraday variability}
\label{sec:wise}

\begin{figure}
\includegraphics[width=\columnwidth]{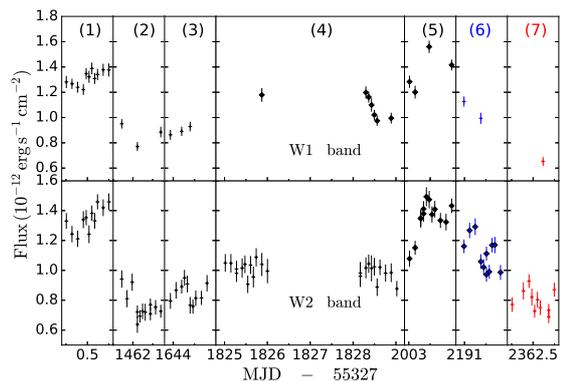}
\caption{$w1$ (upper panel) and $w2$ (lower panel) light curves of J2118$-$0732 obtained from the {\it WISE} data base.
			  The numbers in the upper panel refer to the epoch number of each intraday epoch.
			  The data showing significant intraday variability are differentiated with diamond marks.
			  Epoch (6) in blue and (7) in red correspond to X-2016-05-10 and X-2016-10-27, respectively.}
\label{fig:wise_lc}
\end{figure}

J2118$-$0732 was observed with the {\it WISE}
in four bands $w1$, $w2$, $w3$, $w4$ centred at 3.4, 4.6, 12 and 22\,$\mu$m.
Its long-term light curves of $w1$ and $w2$ bands were constructed from the PSF profile-fit photometric magnitudes
via the near-Earth objects {\it WISE} (NEOWISE) Reactivation Single-exposure data base \citep[][]{2014ApJ...792...30M}
and the AllWISE Multi-epoch Photometry data base (see Fig.~\ref{fig:wise_lc}),
while $w3$ and $w4$ light curves are not available due to the lack of the data in the NEOWISE data base.
Those data with poor signal-to-noise ratios (S/N $<$ 10 or marked as 'null')
or the reduced $\chi^2$ of the profile-fit photometries larger than 2 were excluded.

Significant intraday variability was detected in epoch 4, 5 in the $w1$ band and epoch 5, 6 in the $w2$ band,
with $p$-values $< 0.1$ per cent using the $\chi^2$-test against the null hypothesis of no variation.
This variability time-scale sets an upper limit on the size of the emitting region to 
$\lse 8 \times 10^{-4}$ \,pc, 
which is much smaller than the scale of a putative torus but consistent with that of the jet-emitting region.
The last two epochs of $w1$ and $w2$ light curves coincided with our two \newton\ observations (see
Section~\ref{sec:xmm}) which provided the quasi-simultaneous infrared fluxes together with the X-ray and optical/UV fluxes. 
The average fluxes during these two epochs were calculated to construct the broad-band SEDs in Section~\ref{sec:sed}
and are reported in Table~\ref{tab:opt/uv}. 
A 1$\sigma$ error of 0.2 magnitude was assumed due to the significant intraday variability of the infrared fluxes.

\subsection{Optical/UV photometry and the morphology of the host galaxy}
\label{sec:opt/uv}

\begin{table*} 
	\caption{Infrared/optical/UV photometric results before corrected for Galactic extinction.}
	\label{tab:opt/uv}
	\begin{tabular}{lcccccccccccccccccccccccccccccc} 

		\hline
		         &  \multicolumn{30}{c}{{\it WISE} profile-fit photometric magnitudes}  \\
		Date   &   \multicolumn{15}{c}{$w1$ (3.4 $\mu$m)}    &   \multicolumn{15}{c}{$w2$ (4.6 $\mu$m)}  \\
		 \hline
		2016-05-10  &  \multicolumn{15}{c}{13.529 $ \pm$ 0.2 }  &  \multicolumn{15}{c}{12.516 $ \pm$ 0.2}   \\
		2016-10-27 &  \multicolumn{15}{c}{14.054 $ \pm$ 0.2}    &  \multicolumn{15}{c}{12.876 $ \pm$ 0.2}   \\
		\hline
	                  &  \multicolumn{30}{c}{OM aperture magnitudes}  \\
		 Date       &  \multicolumn{5}{c}{UVW2 (2120 \AA)}     &   \multicolumn{5}{c}{UVM2 (2310 \AA)} & 
		                   \multicolumn{5}{c}{UVW1 (2910 \AA)}    &  \multicolumn{5}{c}{U (3440 \AA)}         & 
		                    \multicolumn{5}{c}{B (4500 \AA)}          &  \multicolumn{5}{c}{V (5430 \AA)}  \\
	 \hline
   2016-05-10        &  \multicolumn{5}{c}{}                          &  \multicolumn{5}{c}{} &  
                                 \multicolumn{5}{c}{19.07 $ \pm$ 0.11} &   \multicolumn{5}{c}{19.34 $ \pm$ 0.08}   &  
                                  \multicolumn{5}{c}{19.82 $ \pm$ 0.05} &   \multicolumn{5}{c}{19.40 $ \pm$ 0.09}   \\
    2016-10-27      &    \multicolumn{5}{c}{}                         &   \multicolumn{5}{c}{} &  
                                  \multicolumn{5}{c}{20.18 $ \pm$  0.17} &  \multicolumn{5}{c}{20.24 $ \pm$ 0.10}    &  
                                   \multicolumn{5}{c}{20.61 $ \pm$ 0.10} &   \multicolumn{5}{c}{}    \\
		\hline
	                        &	 \multicolumn{30}{c}{SDSS PSF-model magnitudes}  \\
		
		 Date           &    \multicolumn{5}{c}{u (3543 \AA)}      &   \multicolumn{5}{c}{g (4770 \AA)} & 
		 						   \multicolumn{5}{c}{r (6231 \AA)}      &  \multicolumn{5}{c}{i (7625 \AA)} & 
		 						    \multicolumn{5}{c}{z (9097 \AA)}  \\
		 \hline
      2000.6741    &  \multicolumn{5}{c}{20.144 $ \pm$  0.056} &  \multicolumn{5}{c}{19.770 $ \pm$  0.019} & 
                                \multicolumn{5}{c}{19.072 $ \pm$  0.023} &  \multicolumn{5}{c}{18.785 $ \pm$  0.021} & 
                                 \multicolumn{5}{c}{18.339 $ \pm$  0.032} \\
         2004.6988 &  \multicolumn{5}{c}{20.101 $ \pm$  0.048} &  \multicolumn{5}{c}{19.691 $ \pm$  0.023} & 
                               \multicolumn{5}{c}{19.118 $ \pm$  0.024} &  \multicolumn{5}{c}{18.892 $ \pm$  0.022} & 
                                \multicolumn{5}{c}{18.399 $ \pm$  0.036} \\
		 2008.7315 &  \multicolumn{5}{c}{19.887 $ \pm$  0.052} &  \multicolumn{5}{c}{19.255 $ \pm$  0.021} & 
		                       \multicolumn{5}{c}{18.601 $ \pm$  0.018} &  \multicolumn{5}{c}{18.274 $ \pm$  0.020} & 
		                        \multicolumn{6}{c}{17.827 $ \pm$  0.029} \\
		\hline
		& \multicolumn{30}{c}{GALEX calibrated magnitudes}  \\
		 Date   &   \multicolumn{15}{c}{FUV (1516 \AA)}    &   \multicolumn{15}{c}{NUV (2267 \AA)}  \\
		 \hline
		 2003-08-06 &  \multicolumn{15}{c}{21.58 $ \pm$ 0.12}  &  \multicolumn{15}{c}{21.38 $ \pm$ 0.10}   \\
		\hline
	\end{tabular}
\end{table*}

\begin{figure}
 \centering
\includegraphics[width=\columnwidth]{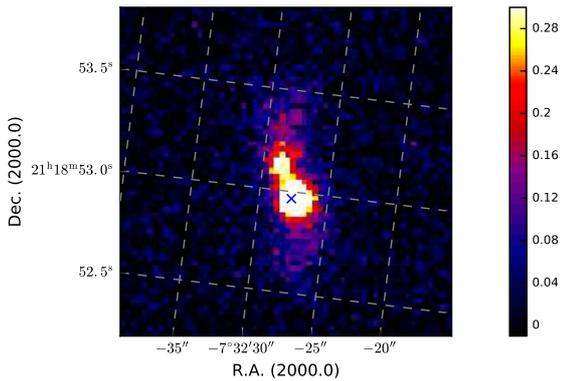}
\caption{SDSS r band image of J2118$-$0732. The '$\times$' mark shows the optical position of the source
and the colour bar is in units of nanomaggy.} 
\label{fig:sdss_image}
\end{figure}

The source was also observed with the OM detector onboard the \newton\ configured in 'imaging' mode
with an exposure of 5\,ks for each filter both on X-2016-05-10 and X-2016-10-27.
We extracted the background-subtracted photometric data using the \texttt{omichain} processing pipeline
with the default parameter settings, as recommended by the \textsc{sas} threads.
The source was not detected in the $UVM2$ and $UVW2$ bands in both observations, and in the $V$ band
in the second observation utilizing the \texttt{omdetect} searching algorithm with a minimum significance of 3. 

We also collected non-simultaneous data from the SDSS Photometric Catalog Data Release 9
\citep[][]{2012ApJS..203...21A} and the Medium Imaging Survey (MIS) observed by
the Galaxy Evolution Explorer \citep[GALEX;][]{2005ApJ...619L...1M}.
The SDSS performed photometric measurements three times on J2118$-$0732 from 2000 to 2008
which showed variations of $\sim$0.5\,mag on time-scale of years.
The source was also detected in the GALEX Far-UV ($FUV$) and Near-UV ($NUV$) bands with magnitudes FUV=21.58
and NUV=21.38 centred at an effective wavelength of 1516\,\AA\ and 2267\,\AA\
in the AB magnitude system\footnote{\url{http://galex.stsci.edu/GalexView/}}.

The collection of the optical/UV magnitudes from the OM, the SDSS, and the GALEX observations
is also reported in Table~\ref{tab:opt/uv}.
All the optical/UV magnitudes were corrected for Galactic extinction using extinction law described in
Section~\ref{sec:spectroscopy} before further analysis. Particularly for the GALEX magnitudes, we used
$A_{\rm FUV}/E(B-V) =$ 8.376 and  $A_{\rm NUV}/E(B-V) =$ 8.741,
following \citet{2005ApJ...619L..15W}.

Of particular interest, the optical image of J2118$-$0732 obtained from the SDSS (see Fig.~\ref{fig:sdss_image}) shows an extended galaxy structure and likely shows a disturbed morphology, possibly suggesting a recent merger.

\section{broad-band Spectral Energy Distribution}
\label{sec:sed}

\begin{figure*} 
  \centering 
  \subfigure[the EC/torus model]{ 
    \includegraphics[width=\columnwidth]{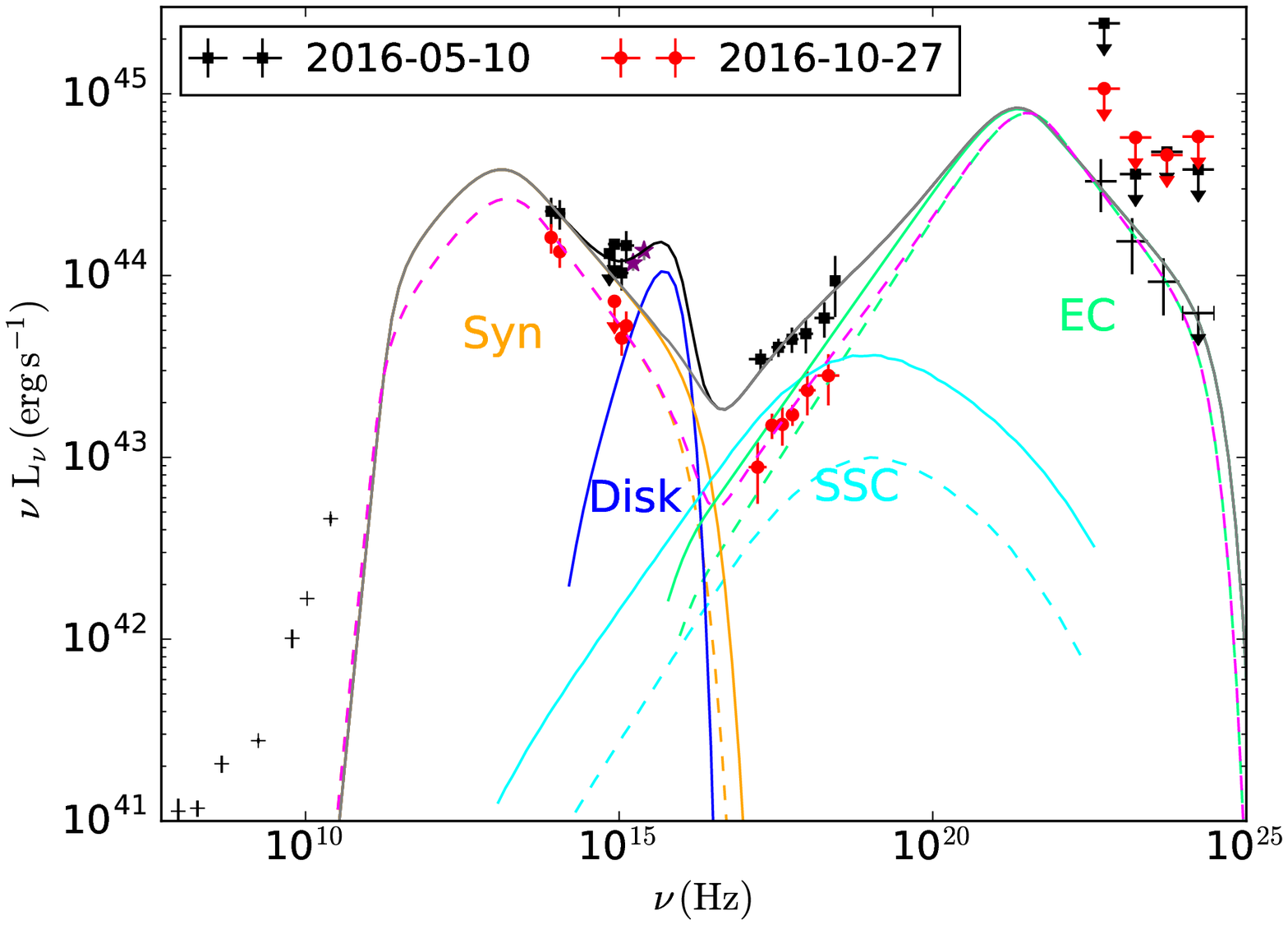} 
  } 
  \subfigure[the EC/BLR model]{ 
    \includegraphics[width=\columnwidth]{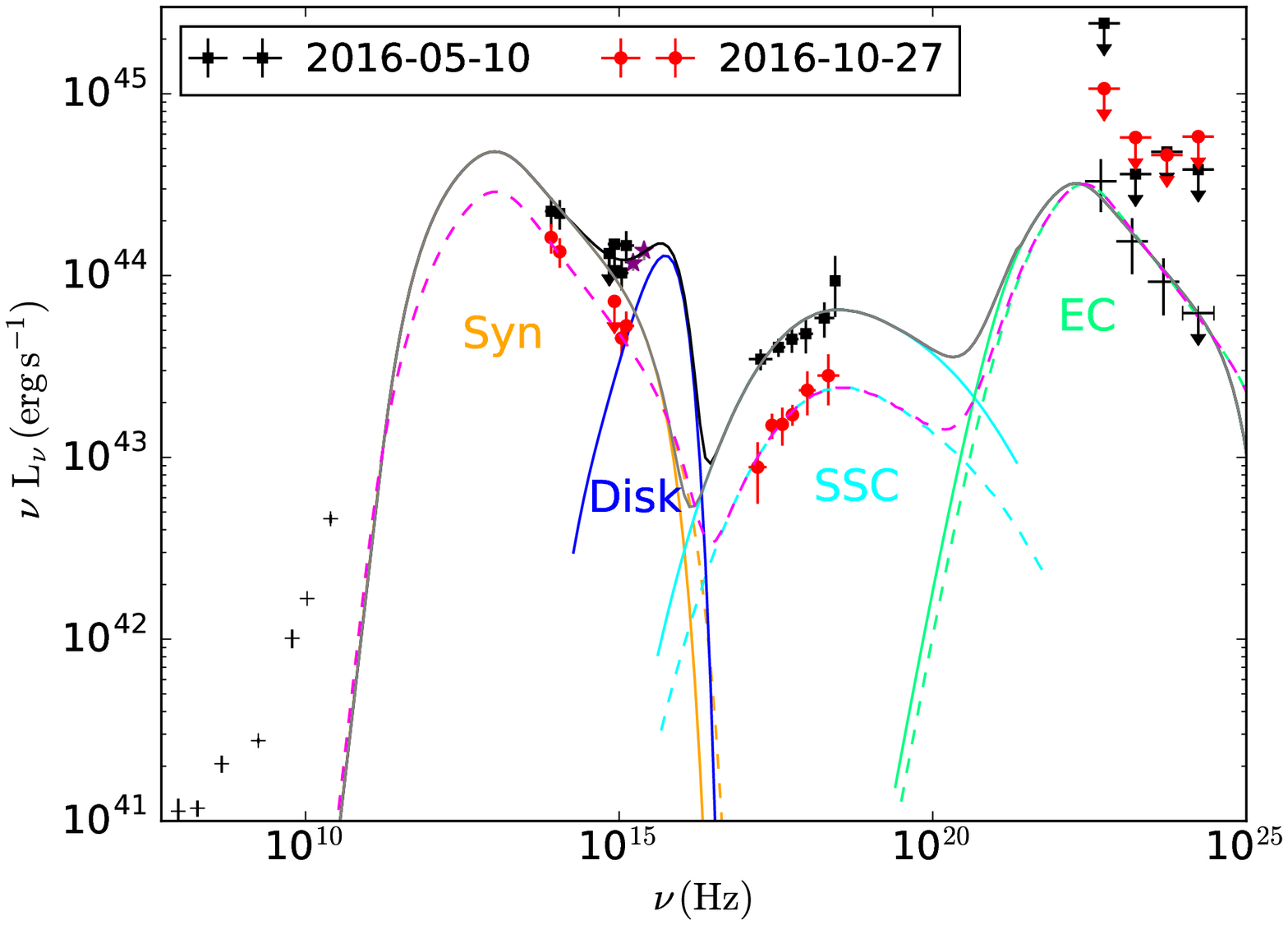} 
  } 
  \caption{The rest-frame broad-band SEDs of J2118$-$0732 and their best-fitting
                 EC/torus models (a) and EC/BLR models (b). 
                 The black squares are the data from 2016-05-10 and the red dots are the data from 2016-10-27
                 covering the infrared, optical/UV, X-ray, and $\gamma$-ray bands.
                 Other non-simultaneous data marked as black crosses are also included into the SED fitting.
                 Besides, the GALEX data denoted as purple stars are also included in the high state
                 to constrain the disc component.
                 The solid grey/black lines show the best-fitting jet/jet$+$disc models for 2016-05-10 data
                 and the dashed magenta lines show the best-fitting jet models for 2016-10-27 data.
                 The solid/dashed orange lines, cyan lines, and spring green lines indicate the synchrotron (Syn),
                 synchrotron-self-Compton (SSC), and External Compton (EC) emissions for the high/low state.
                 And the solid blue lines indicate the disc models for the high state.}
 
  \label{fig:bbsed}
\end{figure*}

\begin{table}
	\caption{Parameters of the best-fitting SED models in the EC/torus and EC/BLR cases.}
	\label{tab:fit_para}
		\begin{threeparttable}
	\begin{tabular}{lcccc} 
	\hline	
				     			& 		  \multicolumn{2}{c}{2016-05-10 (high state)} &\multicolumn{2}{c} {2016-10-27 (low state)} \\
		
		     					                          &  EC/torus  & EC/BLR      &  EC/torus  & EC/BLR   \\
		\hline
		$\gamma_{\rm min}^a$     &    1                    &  78                     &      1                    &      78      \\
		
		$\gamma_{\rm b}^b$    &   575 $\pm$ 146     &  269 $\pm$ 81    &  617 $\pm$ 185  &  372 $\pm$ 111   \\
		
       $\gamma_{\rm max}^c$      &  20000                &   6000                &  15000                &      10000      \\
		
		$p_{1}^d$                             &    2.02               & 1.54                   &  1.92                    &     1.2       \\
		
		$p_{2}^e$                            &    3.8                  & 3.8                      &  3.92                    &      3.9      \\
		
		$N_{0}^f$                            &   120 $\pm$ 61   & 28 $\pm$ 8  & 19 $\pm$ 11 &    2.6   $\pm$ 0.7    \\
	    
		 $\delta^g$                         	 & 	  8.9 $\pm$ 1.5 &	5.9 $\pm$ 0.7    &	 10.3 $\pm$ 1.5  &    5.8 $\pm$ 0.7     \\
		
           $B^h$                                &  1.0 $\pm$ 0.3 &	 3.7 $\pm$ 0.8 & 0.9 $\pm$ 0.3 &  2.6 $\pm$ 0.5      \\
      
        ${\chi^2_{\rm red}}^i$        & 	  0.5                        &	 0.7               &   0.3	                 &    0.5      \\
      
          $L_{\rm bol}^j$           &   2.46     &          2.88       &    & \\
		
		 $\lambda^k$                           & 	 0.06 &		0.07	   &  &         \\
		\hline	
	\end{tabular}
		\begin{tablenotes}
		\item[] 
		$^a$ minimum Lorentz factor of the injected electrons. \\
		$^b$ break Lorentz factor of the injected electrons. \\
		$^c$ maximum Lorentz factor of the injected electrons. \\
		$^d$ low-energy electron spectral index. \\
		$^e$ high-energy electron spectral index. \\
	 	$^f$ electron density parameter in units of 10$^{2}$\,cm$^{-3}$. \\
		$^g$ Doppler boosting factor which equals to bulk Lorentz factor assumed in \citet{2015ApJ...807...51Z}.\\
		$^h$ magnetic field in units of Gauss. \\
		$^i$ reduced $\chi^2$ of the SED fittings with one-zone leptonic jet model. \\
		$^j$ bolometric luminosity  $L_\mathrm{bol}$ = $L_\mathrm{disc}$ in units of 10$^{44}$\, erg\,s$^{-1}$. \\
		$^k$ Eddington ratio calculated from $\lambda = L_\mathrm{bol}/L_\mathrm{Edd}$.
		\end{tablenotes}
		\end{threeparttable}
\end{table}

For $\gamma$-ray loud NLS1s like J2118$-$0732, which exhibit the hybrid properties of both NLS1s and blazars, it would be revealing to study their broad-band SEDs. 
Here, we made use of the \newton\ data and the {\it WISE} data to construct the quasi-simultaneous SEDs
in a relatively high state (on 2016 May 10) and low state (on 2016 October 27). 
The radio spectrum was taken from the NED and VizieR websites as presented in Section~\ref{sec:radio}.
We also made use of the GALEX data in the $UV$ band to constrain the disc component in the high state because of the similar UV flux levels of the GALEX data and the OM data.
The broad-band SEDs are shown in Fig.~\ref{fig:bbsed}.
The source was not detected in $\gamma$-rays at the epochs around the \newton\ observations within a time span of 60 d (see the study on the $\gamma$-ray light curve in Section~\ref{sec:fermi-results}).
The flux limits derived above are denoted as arrows.  
As a first-order approximation, the average $\gamma$-ray spectrum obtained from all the data over the total 9-yr time span was used instead, which is consistent with the upper limits.
We assume that the $\gamma$-ray emission in the SED of J2118$-$0732 at the epochs concerned can be approximated by the averaged spectrum. 
This allows us to perform spectral modelling of the SED, which is expected to be still revealing, though the best-fitting model and parameters may suffer from uncertainties to some extent due to possible spectral variations. 

To fit the broad-band SED, we first used a simple one-zone leptonic jet model, which consists of
synchrotron, synchrotron self-Compton (SSC), and external-Compton (EC) processes
\citep[e.g.][]{2009ApJ...707L.142A,2012MNRAS.426..317D,2013ApJ...768...52P,2015ApJ...798...43S}.
The $\chi^2$ minimization technique is applied to perform the fitting.
The radio emission may come from radio lobes at larger scales and/or superposition of multiple jet components
and thus does not fit to the synchrotron component which is self-absorbed below 10$^{11}$\,Hz. 
Two scenarios were considered for the EC scattering process:
a seed photon field dominated from the torus and from the BLR, respectively.
The energy density of the former was assumed to be $3 \times\,10^{-4}$\,erg\,cm$^{-3}$
\citep[see more details in][]{2007ApJ...660..117C, 2008MNRAS.387.1669G} and the latter
was estimated from the broad \hb\ line luminosity derived in Section~\ref{sec:spectroscopy}
\citep[see more details in][]{2015ApJ...798...43S,2015ApJ...807...51Z}.
The energy distribution of the injected relativistic electrons was assumed to be a broken PL in the range of [$\gamma_{\rm min}$, $\gamma_{\rm max}$] with a break energy $\gamma_{\rm b}$.
The value of $\gamma_{\rm min}$ was taken as $\gamma_{\rm min} =  1$ in the EC/torus model
which was constrained from the soft X-ray slope of the SSC bump and $\gamma_{\rm min} = 78$ in the EC/BLR model
which is the mean value of the $\gamma_{\rm min}$ distribution for a GeV-NLS1 sample \citep[][]{2015ApJ...798...43S}.
The $\gamma_{\rm max}$ is usually poorly constrained and
does not significantly affect our results, so it was fixed at a large value. 
The two indices $p_1$,  $p_2$ of the broken PL were derived from the spectral indices of the observed SEDs.

Given the prominent broad emission lines in the optical spectrum of J2118$-$0732,
a thermal emission contributed by an accretion disc is very likely present in J2118$-$0732.
This is particularly true when J2118$-$0732 was in the high state, where the UV fluxes calculated
from the OM data and the GALEX data started to rise towards the frequency range near the peak of the big blue bump.
We used a multitemperature blackbody model of a standard thin disc to account for this component in the high state.
But for the low state, we did not add a disc model due to the lack of available data points
covering the frequency range where the disc is the strongest.
A black hole mass of 3.4 $\times$ 10$^7$\,$M_{\sun}$ was employed.
We also note that the optical fluxes should be treated as upper limits (marked as arrows in Fig.~\ref{fig:bbsed})
caused by the contamination from the host galaxy, as revealed by the SDSS optical image (see Fig.~\ref{fig:sdss_image}) which shows an
extended galaxy structure, while the UV fluxes\footnote{\label{foot:uvbands}The observational $U$ band of the OM will be
blueshifted to the UV band at the rest frame considering the redshift of J2118$-$0732.} are not. 
Since the hard X-ray spectrum is more flat than the average of radio-quiet NLS1s,
we assume that it is dominated by jet emission (in addition to fainter emission from the corona).
For simplicity, in the SED modelling, we therefore do not include coronal emission.

Both the EC/torus and the EC/BLR models can adequately fit the broad-band SEDs and 
the disc component can be roughly constrained in the high state using the additional GALEX data.
The parameters of the fitting are given in Table~\ref{tab:fit_para}, together with their 1$\sigma$ errors.
As can be seen in Fig.~\ref{fig:bbsed}, the infrared to ultraviolet is dominated by the synchrotron radiation
and the $\gamma$-ray band is dominated by the EC radiation, respectively.
The EC peak of the EC/torus model is broader than that of the EC/BLR model and extends to the X-ray band
while only the SSC peak dominates the X-ray band in the EC/BLR model.
The best-fitting parameters show that the EC/BLR model contains a higher magnetic field and
the Doppler boosting factor is smaller compared with that of the EC/torus model.
Noticing that the main difference of the broad-band SEDs of these two models lies in the luminosity from 10\,keV up to
100\,MeV, hard X-ray and/or soft $\gamma$-ray data would help to distinguish between the EC/torus and the EC/BLR models.
However, only the upper limit of the source in the hard X-ray and/or soft $\gamma$-ray band can
be constrained from surveys like the latest 105-month Swift-BAT all-sky hard X-ray survey in the 14--195\,keV
\citep[][]{2018ApJS..235....4O} which is not sensitive enough to discriminate between the two models for the SED yet.
Further data are needed to distinguish between the two models and give better understandings of the EC process in astrophysical jets. 

The multiwavelength SEDs of J2118$-$0732 changed from a high state on 2016 May 10 to
a low state on 2016 October 27 with a synchronous drop of the broad-band fluxes from infrared to X-rays in five months.
The electron density parameter varied by about an order of magnitude between the two states for both models.
The Eddington ratio was estimated to be $\approx0.07$ in the high state.
For the low state, a reliable estimate of the Eddington ratio cannot be made,
but it is probably lower than that of the high state, as indicated by its lower UV fluxes.

The best-fitting parameter values, including the magnetic fields, the Doppler boosting factors, and the electron densities are within the typical range of some other GeV NLS1s
\citep[e.g.][]{2009ApJ...707L.142A,2011MNRAS.413.1671F,2012MNRAS.426..317D,2013ApJ...768...52P,2015ApJ...798...43S,2015AJ....150...23Y}
and similar to those of blazars \citep[e.g.][]{2010MNRAS.401.1570T,2015ApJ...807...51Z}.
However, the exact values of these parameters may be subject to relatively large uncertainties, given the degeneracies among some of the parameters \citep[e.g.][]{1998ApJ...509..608T,2012ApJ...752..157Z} as well as the non-simultaneity of the multiwavelength data.

\section{Discussions}
\label{sec:discussion}

\subsection{Blazar-like properties of J2118$-$0732}
\label{sec:blazar-like}

Blazar-like characteristics have been found in some RL NLS1s \citep{2008ApJ...685..801Y},
suggesting the presence of relativistic jets which was later confirmed by
the detection of $\gamma$-ray emission from some of these objects with the {\it Fermi}. 
The similar multiwavelength properties of J2118$-$0732 make it a new member of this interesting and rare class of objects.
The flat radio spectrum above 1.4\,GHz of J2118$-$0732 is probably the result of a superposition of several jet
components \citep{1981ApJ...243..700K} while the steep spectrum below 1.4\,GHz is likely coming from radio lobes.
The short time-scale of infrared emission is consistent with the size of the jet-emission region.
Besides, the hard X-ray spectrum of J2118$-$0732 is similar to those of some blazar-like NLS1s,
hinting perhaps at a jet origin.
Furthermore, \citet{2014ApJS..215...14D} found that the {\it WISE} mid-infrared colours of J2118$-$0732 are similar to those
of the confirmed $\gamma$-ray emitting blazars that occupy the so-called {\it WISE} blazar locus
\citep[][]{2013ApJS..206...12D} in the {\it WISE} colour space,
which makes J2118$-$0732 a '{\it WISE} blazar-like radio-loud source'. 

The blazar-like characteristics of J2118$-$0732 are also reminiscent of another $\gamma$-ray-emitting NLS1s -- PKS~2004$-$447.
PKS~2004$-$447 has a very high radio loudness (1710 $< R <$ 6320) and a low BH mass 
($M_{\rm BH} = 10^{6.7}\,M_{\sun}$) at a redshift of 0.24.
The jet emission is also suggested to dominate at multiple wavelengths in PKS~2004$-$447 which is supported by the
multiwavelength analysis and the SED modelling
\citep{2009ApJ...707L.142A,2013ApJ...768...52P,2015MNRAS.453.4037O}.
Moreover, both J2118$-$0732 and PKS~2004$-$447 show weak \feii\ emission 
and lack an obvious soft X-ray excess\footnote{We note a tentative soft excess
has been reported in only one of the three \newton\ observations of PKS~2004$-$447
\citep[][]{2006MNRAS.370..245G,2016A&A...585A..91K}.}, while a soft X-ray excess has been detected in most
other $\gamma$-loud NLS1s that have good-quality X-ray spectra
\citep[e.g.][]{2008MNRAS.388L..54D,2014MNRAS.438.3521D,2015A&A...575A..13F,2018MNRAS.475..404K,2018MNRAS.476...43L}.

\subsection{Association of the $\gamma$-ray source with J2118$-$0732}
\label{sec:association}

Even though the blazar-like characteristics make the RL NLS1 J2118$-$0732 a highly possible counterpart for the
$\gamma$-ray source detected by the {\it Fermi}-LAT and both \citet{2018ApJ...853L...2P}
and FL8Y reported a high association probability of the source, we still make a sanity check of the association
between the two within the relatively large error circle of the $\gamma$-ray source detection. 

There are 11 radio sources detected in the FIRST survey within the gamma-error circle above the 1.5\,mJy flux threshold
and J2118$-$0732 is the brightest one with its radio flux higher than the combined flux of the rest 10 sources. 
Its monochromatic radio luminosity ($2.7 \times 10^{41}$\,erg\,s$^{-1}$) at 1.4\,GHz
and the 9-yr averaged monochromatic $\gamma$-ray luminosity at 1\,GeV ($0.9 \times 10^{44}$\,erg\,s$^{-1}$) of the $\gamma$-ray source
follow the correlation between the two luminosities as claimed by \citet{2016ApJS..226...20F}.
The source detection algorithm on the EPIC images found 45 X-ray sources within the gamma-error circle and
J2118$-$0732 is the brightest X-ray source with its X-ray flux in accordance with other $\gamma$-ray emitting {\it Fermi} blazars from a sample in \citet{2016ApJS..226...20F}.
Although a small part of the $\gamma$-ray error circle is not covered by the EPIC field of view,
the only detection of J2118$-$0732 within this error circle by the RASS can set an upper limit on the fluxes of other X-ray sources. 
Furthermore, J2118$-$0732 is the only X-ray bright source in the EPIC image which has a radio counterpart among 11
radio sources detected in the FIRST catalogue, making it most likely associated with the $\gamma$-ray source.
No known blazar is discovered by checking every X-ray source and radio source found above. 
However, an unambiguous and independent confirmation should come from future monitoring observations
which search for the correlated variability of $\gamma$-ray emission with emission from other wavelengths.

\subsection{$\gamma$-ray Variability}
\label{sec:gamma_vaviability}

So far, only a limited number of $\gamma$-loud NLS1s have been detected by {\it Fermi}-LAT including J2118$-$0732.
However, the $\gamma$-ray properties are different from source to source.
Three objects, PMN~J0948$+$0022, SBS~0846$+$513, and 1H~0323$+$342, have shown short-term strong
$\gamma$-ray flares emerging and dying out within one month \citep[][]{2012MNRAS.426..317D,2015MNRAS.446.2456D}. 
Other $\gamma$-ray sources have not shown strong flares so far.

J2118$-$0732 was in a relatively high-flux state during 2009--2013 
and dimmed below the detection limit of {\it Fermi}-LAT after 2013.
During the active 4 yr, at least one flare was detected.
The long-term activity with relatively high fluxes lasting for years of J2118$-$0732 may be a common feature
since similar behaviours have been observed among some other $\gamma$-ray loud NLS1s: a nine-month flare
was observed in a long time variability analysis of FBQS~J1644$+$2619 \citep{2015MNRAS.452..520D},
and a relatively high-flux state was also observed in SDSS~J144246.29$+$474129.4
in late 2012 lasting for several months \citep{2015arXiv151005584L}.
This may indicate that a number of RL NLS1s may radiate in $\gamma$-rays
but remain undetected by the {\it Fermi}-LAT because they spend more of the time in a low-flux state. 
Therefore, because of this characteristic of long-term variability, more RL NLS1s, especially those with the highest radio loudness, are expected to be observed in the $\gamma$-ray band in the future.

\section{Summary}
\label{sec:summary}

$\gamma$-ray-emitting NLS1s have been providing new insights into the formation and evolution of relativistic jets
and the jet-disc coupling mechanism under the extreme conditions of low BH masses and high accretion rates.
Here, we present our discovery of a new $\gamma$-ray-emitting NLS1-- SDSS~J211852.96$-$073227.5
by analysing the SDSS spectrum, the {\it Fermi}-LAT, and the \newton\ observational data.
The multiwavelength observations presented here allow a comprehensive study of the properties of
its broad-band radiation. The main results are summarised below:

\begin{enumerate}

\item[1.] The optical spectrum of J2118$-$0732 from the SDSS confirms its NLS1 nature:  a broad \hb\ emission line with a width (FWHM) of
1585\,km\,s$^{-1}$ and a flux ratio \oiii/\hb\ $\approx1.7$ even though the \feii\ emission is rather weak.
The estimated BH mass is $\sim$3.4 $\times 10^7\,M_{\sun}$ and the Eddington ratio is $\sim$0.15.

\item[2.] The {\it Fermi}-LAT observations centred at J2118$-$0732 detected a new $\gamma$-ray source
with a 9-yr averaged isotropic luminosity of 0.7 \,$\times\,10^{45}$\,erg\,s$^{-1}$
in the 0.1--300\,GeV rest-frame energy band.
The new $\gamma$-ray source experienced a relatively high state for about 4\,yr
and remained in a quiescent state in other {\it Fermi}-LAT observations.
At least one flare was emerging during the long-term activity.

\item[3] Our two \newton\ observations covering a significant part of the $\gamma$-ray source error circle
reveal that J2118$-$0732 is the brightest X-ray source in the field, and the one with large X-ray variability.
The X-ray spectrum of J2118$-$0732 is flat, with no significant evidence of a soft excess.

\item[4.] Other blazar-like properties of J2118$-$0732 include a large radio loudness with a flat/inverted radio
spectrum and remarkably rapid infrared variability of less than a day, detected by {\it WISE}.
This short time-scale implies a compact emission region much more compact than the torus.

\item[5.] The broad-band SED can be modelled by emission from a simple one-zone leptonic jet, and the flux variation from infrared to X-rays in five months can be explained by changes of the jet parameters.
Besides, there may exist a relatively strong disc component, at least in the high state.  
We note the caveat that, however, the exact values of the fitted parameters in our model may be subject to relatively large uncertainties, given the degeneracies among some of the model parameters as well as the non-simultaneity of the multiwavelength data.

\end{enumerate}

\section*{Acknowledgements}

We thank the anonymous referee for helpful comments and suggestions.
This work is supported by the Natural Science Foundation of China (grants No. 11473035, 11573034, and 11773037). SY and YL acknowledge the support by the KIAA-CAS Fellowship, which is jointly supported by Peking University and Chinese Academy of Sciences.
SY thanks for the support by Boya fellowship. 
Funding for SDSS-III has been provided by the Alfred P. Sloan Foundation, the Participating Institutions,
the National Science Foundation, and the U.S. Department of Energy Office of Science.
The SDSS-III web site is \url{http://www.sdss3.org/}.
SDSS-III is managed by the Astrophysical Research Consortium for the Participating Institutions of the SDSS-III
Collaboration including the University of Arizona, the Brazilian Participation Group, Brookhaven National Laboratory,
Carnegie Mellon University, University of Florida, the French Participation Group, the German Participation Group,
Harvard University, the Instituto de Astrofisica de Canarias, the Michigan State/Notre Dame/JINA Participation Group,
Johns Hopkins University, Lawrence Berkeley National Laboratory, Max Planck Institute for Astrophysics,
Max Planck Institute for Extraterrestrial Physics, New Mexico State University, New York University, Ohio State University,
Pennsylvania State University, University of Portsmouth, Princeton University, the Spanish Participation Group,
University of Tokyo, University of Utah, Vanderbilt University, University of Virginia, University of Washington,
and Yale University.
The {\it Fermi}-LAT Collaboration acknowledges generous ongoing support from a number of agencies and institutes
that have supported both the development and the operation of the LAT as well as scientific data analysis.
These include the National Aeronautics and Space Administration and the Department of Energy in the United States,
the Commissariat \`{a} l'Energie Atomique and the Centre National de la Recherche Scientifique / Institut National de
Physique Nucl\'{e}aire et de Physique des Particules in France, the Agenzia Spaziale Italiana and the Istituto Nazionale di
Fisica Nucleare in Italy, the Ministry of Education, Culture, Sports, Science and Technology (MEXT), High Energy
Accelerator Research Organization (KEK) and Japan Aerospace Exploration Agency (JAXA) in Japan, and the K. A.
Wallenberg Foundation, the Swedish Research Council and the Swedish National Space Board in Sweden.
Additional support for science analysis during the operations phase is gratefully acknowledged from the Istituto
Nazionale di Astrofisica in Italy and the Centre National d'Etudes Spatiales in France.
Based on observations obtained with XMM-Newton, an ESA science mission with instruments and contributions directly funded by ESA Member States and NASA. 
This publication makes use of data products from the Wide-field Infrared Survey Explorer,
which is a joint project of the University of California, Los Angeles, and the Jet Propulsion Laboratory/California Institute
of Technology, funded by the National Aeronautics and Space Administration.
This work is based in part on observations made with the Galaxy Evolution Explorer (GALEX).
GALEX is a NASA Small Explorer, whose mission was developed in cooperation with the Centre National d'Etudes
Spatiales (CNES) of France and the Korean Ministry of Science and Technology. GALEX is operated for NASA by the
California Institute of Technology under NASA contract NAS5-98034. 
This research has made use of the NASA/IPAC Infrared Science Archive and Extragalactic Database (NED),
which are operated by the Jet Propulsion Laboratory, California Institute of Technology, under contract with the National Aeronautics and Space Administration. 
This research has also made use of the VizieR catalogue access tool, CDS, Strasbourg, France.









\bsp	
\label{lastpage}
\end{document}